\documentclass[runningheads]{llncs}

\usepackage{cite}
\usepackage{amsmath,amssymb,amsfonts}
\usepackage{graphicx}
\usepackage{textcomp}
\usepackage{xcolor}
\usepackage{algorithmic}
\usepackage{algorithm}
\usepackage{marvosym}

\begin{document}


\title{Multi-dimensional Data Quick Query for Blockchain-based Federated Learning}

%
%


\author{Jiaxi Yang\inst{1,2} \and
Sheng Cao \inst{1,2}\textsuperscript{(\Letter)}\and
Peng Xiangli\inst{3} \and{Xiong Li}\inst{1,2} \and{Xiaosong Zhang}\inst{1,2} }
\authorrunning{J. Yang et al.}
\titlerunning{Multi-dimensional Data Quick Query for Blockchain-based FL}
%
\institute{Shenzhen Institute for Advanced Study, University of Electronic Science and Technology of China, Shenzhen, China \and
School of Computer Science and Engineering, University of Electronic Science and Technology of China, Chengdu, China\\ 
\and The Fifth Electronic Research Institute of MIIT, Guangzhou, China 
}

\maketitle

\begin{abstract}
Due to the drawbacks of Federated Learning (FL) such as vulnerability of a single central server, centralized federated learning is shifting to decentralized federated learning, a paradigm which takes the advantages of blockchain. A key enabler for adoption of blockchain-based federated learning is how to select suitable participants to train models collaboratively. Selecting participants by storing and querying the metadata of data owners on blockchain could ensure the reliability of selected data owners, which is helpful to obtain high-quality models in FL. However, querying multi-dimensional metadata on blockchain needs to traverse every transaction in each block, making the query time-consuming. An efficient query method for multi-dimensional metadata in the blockchain for selecting participants in FL is absent and challenging. In this paper, we propose a novel data structure to improve the query efficiency within each block named MerkleRB-Tree. In detail, we leverage Minimal Bounding Rectangle(MBR) and bloom-filters for the query process of multi-dimensional continuous-valued attributes and discrete-valued attributes respectively. Furthermore, we migrate the idea of the skip list along with an MBR and a bloom filter at the head of each block to enhance the query efficiency for inter-blocks. The performance analysis and extensive evaluation results on the benchmark dataset demonstrate the superiority of our method in blockchain-based FL.

\end{abstract}
\section{Introduction}

As a special distrbuted machine learning framework, FL, in which allows multiple data owners to train machine learning models collaboratively with their data stored locally, is much popular in the present age \cite{yang2019federated}. However, centralized FL still faces some challenges such as the failure of a single central server, etc. With the overwhelming development of blockchain technology, it is possible to leverage some advantages of blockchain to FL and construct a decentralized FL paradigm named blockchain-based FL \cite{kim2019blockchained, warnat2021swarm}. In blockchain-based FL, blockchain is able to enhance the robustness, trust, security of FL, as well as providing a credible cooperation mechanism among participants.

When the aggregation server initializes a FL task, it need to select a set of data owners to participate. Selecting participating nodes according to their data type without knowing the metadata information of data owners is challengeable. Through providing secure data storage platform in blockchain-based FL, data owners can announce the description of their data called metadata in the community via blockchain\cite{zhang2020blockchain}. When the metadata is queried on the blockchain, the aggregation server can abtain the candidate participants list from nearest proxy server and invite these nodes to participate in FL\cite{kumar2021blockchain}. And we will introduce the process in section 3.


However, on the one hand, the query efficiency on the existing blockchain is extremely low \cite{peng2019vql}. With the increasing number of data owners registering, it cannot meet the large query requirements of the aggregation server when selecting nodes. On the other hand, in the real scenario of choosing data owners in FL, the query condition may usually be composed by multi-dimensional continuous-valued attributes and discrete-valued attributes \cite{zhang2020blockchain}. Existing query methods on the blockchain can only cater for single-dimensional hash value. It is inefficient to store multi-dimensional attributes in the single dimensional data structures, since it needs to do intersectional operation when we need to query for multi-dimensional attributes \cite{theodoridis1996spatio}. For multi-dimensional query condition with both continuous-valued attributes and discrete-valued attributes on the blockchain, yet there is no appropriate query method to satisfy this kind of query demand \cite{zhang2019gem}. In this paper, we propose a method for the query process of both inter-block and intra-block. Our contributions are listed in the following points:


\begin{itemize}
	\item We formulate the selection of participating nodes in blockchain-based FL as the metadata query problem on blockchain. We divide this query problem into intra-block query and inter-block query and put forward schemes for them respectively. 
	\item For intra-block query, we modify the structure of the block and construct an MerkleRB-Tree in each block. Query schemes for both discrete-valued attributes and continuous-valued attributes are proposed. 
	\item For inter-block query, we apply the skip list and implement an inter-block query scheme with bloom-filters and MBR for discrete-valued attributes and continuous-valued attributes respectively.
	\item We analysis the performance of the query schemes we propose. The results of the comparative experiments with the baseline method show our schemes are more efficient.
\end{itemize}

\begin{figure*}[t!]
  	\centering{\includegraphics[scale = 0.43]{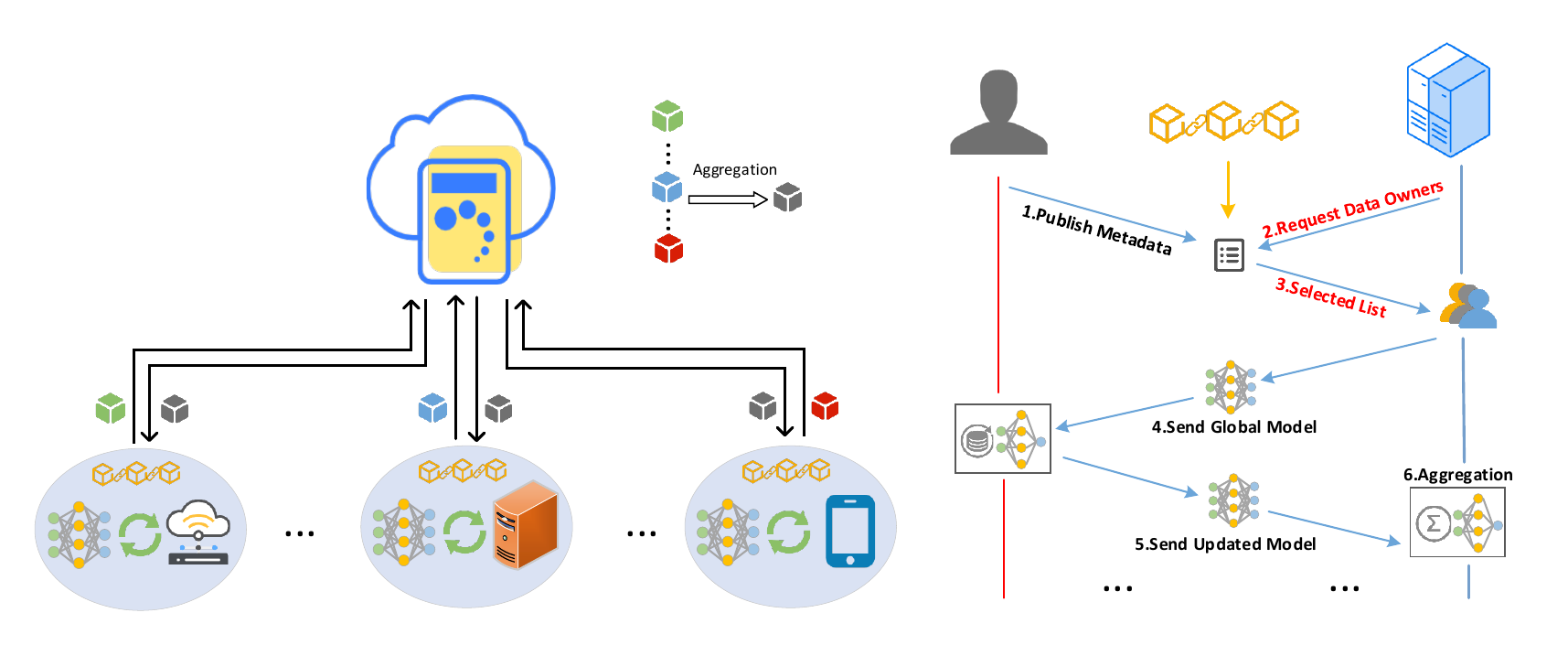}}
	\caption{Blockchain-based FL Framework}
	\label{framework}
\end{figure*}

\section{Related Work}

\subsection{Blockchain Empowered Federated Learning}
Since tranditional centralized FL faces a number of challenges~\cite{kairouz2019advances} such as lack of a secure and credible cooperation mechanism etc, an increasing number of studies focus on empowering FL with blockchain \cite{billah2022systematic}. Being empowered with blockchain, FL owns a credible incentive and contribution measurement mechanism as well as strengths its security\cite{2020Blockchained}.




Besides, blockchain provides a trusted storage mechanism for FL, allowing data to be shared securely. Data owners can leverage blockchain to publish their metadata information and then aggregation servers can select participating nodes by querying the metadata on the blockchain according to the data type \cite{kumar2021blockchain}. Zhang et al. propose a FL protocol based on blockchain in which the nearest proxy server can help to query metadata on blockchain and return the set of selected participating nodes~\cite{zhang2020blockchain}. However, these studies do not focus on the query efficiency of metadata in blockchain-based FL, nor did they change the original block structure.



\subsection{Query on the blockchain}
For query on the blockchain, traversing every transaction of each block can be regarded as time-consuming.
Current studies show that using external databases can improve the query efficiency of blockchain query \cite{nathan2019blockchain, muzammal2019chainsql}. By establishing an efficient query layer, EtherQL, Li et al. propose a quick query method that imports block data into an off-chain database using the Ethereum listening interface~\cite{li2017etherql}. Peng et al. propose a three-tier blockchain query architecture, which saves the time to traverse unnecessary blocks~\cite{peng2019vql}. Zhang et al. design new data structures named $Gem^2 Tree$ which can be effectively maintained by blockchain, significantly reducing the storage and computing cost of smart contract~\cite{zhang2019gem}. However, these schemes are hardcoded and cannot be well adapted to different query conditions and do not consider the problem of the inter-block query.


\section{Problem Formulation}
\subsection{System Framework}
In the training process of blockchain-based FL, the data owners register the FL community and publish the metadata to the blockchain. It is noticeable that the metadata generally refer to the description of the data type of the data owners. When the metadata is queried by the aggregation server, it can get the candidates list according to the task requirements. Then the aggregation server initializes the machine learning model and allocates it to participants for local training. Finally, after getting the updated models from participants, the aggregation server aggregates them to update the global model. The system paradigm is detailed in Fig. \ref{framework}. 






 
\subsection{Query Metadata on the Blockchain}
In our system framework, the aggregation are responsible for querying the metadata on the blockchain. Moreover, when the aggregation servers select parties to participate in the FL task, they usually select different parties to join in based their type of data sets according to the requirements of machine learning model training task. In the metadata, there are discrete-valued attributes and also continuous-valued attributes. Each query condition may contain multi-dimensional discrete-valued attributes and continuous-valued attributes. Therefore, we can regard the problem as the mixed multi-dimensional query of continuous-valued attributes and discrete-valued attributes.

However, it is inefficient to use existing methods to solve this problem. In the traditional way, in order to find the data owners who have this type of data set, they may need to traverse all the transactions in every block and check whether each query condition is satisfied. It is time-consuming if we traverse all the transactions in the blockchain. If the query condition of continuous-valued attributes is multi-dimensional, it will increase the difficulty and cost of querying to a greater extent~\cite{theodoridis1996spatio}. Therefore, the problem is how to design a data structure in the blockchain to query both discrete-valued attributes and multi-dimensional continuous-valued attributes more efficiently. In the next two sections, we divide this problem into the intra-block query and inter-block query and describe the solution we propose for this problem respectively.


\section{Intra-block Query Scheme}

\subsection{The structure of MerkleRB-Tree}
\begin{figure*}[t!]
	\centering{\includegraphics[scale = 0.38]{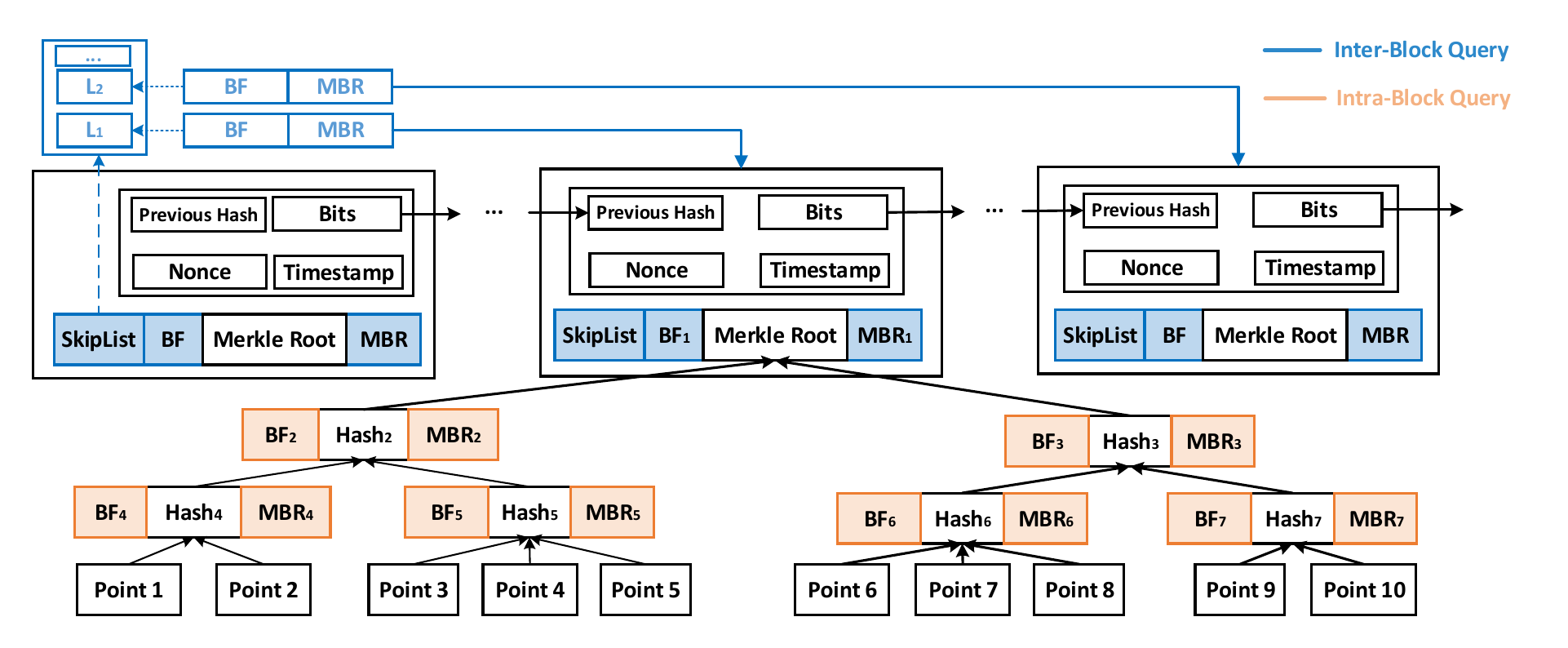}}
	\caption{The query structure for both discrete-valued and continuous-valued attributes on blockchain}
	\label{arch}
\end{figure*}

In order to make the intra-block query quicker, we apply the idea of MR-Tree \cite{yang2008spatial} and bloom filter together to construct a new structure named MerkleRB-Tree. MerkleRB-Tree extends the advantages of MR-Tree and bloom filter. We use it for multi-dimensional query of both continuous-valued and discrete-valued attributes. As shown in Figure \ref{arch}, MerkleRB-Tree verifies the integrity of the whole tree based on Merkle Hash Tree. Each internal node contains a hash value, a bloom filter, and an MBR \cite{kamel1993hilbert}. The bloom filter in each node can be used to check whether there are existing transactions in the current subtree that satisfy the discrete-value query condition. MBR covers the range of continuous-valued attributes of all transactions of every dimensions in its subtree.


\subsection{Intra-block Query of Continuous-valued Attributes}
For querying multi-dimensional continuous-valued attributes, it is inefficient to take the intersection after querying the multiple dimensions separately for querying multi-dimensional continuous-valued attributes. In this section, we focus on the query of multi-dimensional continuous-value. In the MerkleRB-Tree, the spatial range of the MBR at the root node of the whole tree. In the query process, we use the recursion method to traverse all the child nodes of the current node. If there is an intersection between the spatial scope of the multi-dimensional query condition and the child node, then we continue to search down the current subtree. The specific algorithm is shown in Algorithm 1. Using the above method for multi-dimensional range query, we can save the time cost of traversing unnecessary nodes in MerkleRB-Tree and improve the efficiency of the query process.

\begin{algorithm}[!b]  
\caption{IntraQuery($Node, MBR_{query}$)}
\hspace*{0.02in}{\bf Input:} Query condition $q = <Q_{discrete}, MBR_{query}>$, query tree $Tree^{}$\\
\hspace*{0.02in}{\bf Output:} result $\Omega$
\begin{algorithmic}[1]
 \IF{$Node$.isLeaf()=False}
    \FOR{$MBR_{child}^i$ for $Node$}
        \IF{$MBR_{child}^i$ intersects $MBR_{query}$ AND \\$BF_{child}^i$.isContain($Q_{discrete}$)}
            \STATE $IntraQuery(Node.Child(i), MBR_{query})$
        \ENDIF
    \ENDFOR
\ELSE
    \STATE Add this to the result set $\Omega$
\ENDIF
\end{algorithmic}
\end{algorithm}

\subsection{Intra-block Query of Discrete-valued Attributes}
For querying discrete-valued attributes, we add a bloom filter~\cite{bloom1970space} in each node of MerkleRB-Tree. A Bloom filter is a long binary vector and a series of random mapping functions that can be used to check whether an element is not in a set. In MerkleRB-Tree, the bloom filter of each node can determine whether all the transactions in the subtree do not satisfy the query condition of discrete-valued attributes. In other words, the non-leaf node's bloom filter is the sum of all its child nodes' bloom-filters, which we represent in formula $(\ref{BFParent})$.
\begin{equation}
	\begin{split}
		BF_{parent} = BF_{child}^1+BF_{child}^2+...BF_{child}^n
	\end{split}
	\label{BFParent}
\end{equation}

For each discrete-valued query condition, we start it from the root node of MerkleRB-tree. For all the transactions in the left and right subtrees that do not satisfy the discrete-valued query condition, bloom-filters in each node are used to find these subtrees and not query them.

\section{Inter-block Query scheme}
\subsection{Inter-block Index structure}
In figure \ref{arch}, an MBR and a bloom filter are added to the block header. For querying discrete-valued attributes, we use the bloom filter at the head of the block to verify whether the block contains any transactions that satisfies the query condition. If no transaction satisfies the query condition, we do not need to query within the block. Similarly, for the inter-block query of the multi-dimensional continuous-valued attributes, we also need to check whether the range space of query condition has an intersection with the MBR at the head of each block.

However, traversing all the blocks in the blockchain is time-consuming. In order to solve this problem, inspired by the idea of dichotomy, we apply a skip list and put forward an efficient inter-block query method. The architecture of the inter-block query is shown in Figure \ref{arch}, and each level of the skip list includes a bloom filter and an MBR denoted as $SkipList_{BF}^i$ and $SkipList_{MBR}^i$. For $SkipList_{BF}^i$ which is used to query discrete-valued attributes, the $i^{th}$ level's bloom filter in the skip list can be used to check whether there are satisfied transactions in the next $\alpha^i$ blocks. We can use formula $(\ref{BFSkipList})$ to represent it. 
\begin{equation}
	BF_{SkipList}^i = BF_{block}^{current}+...+BF_{block}^{current+\alpha^i}
	\label{BFSkipList}
\end{equation}

 Similar to $BF_{SkipList}^i$, the $i^th$ MBR in $MBR_{SkipList}^i$ is the minimum bound rectangle of all the MBRs at each block's head which we represent in formula $(\ref{MBRSkipList})$.

\begin{equation}
	MBR_{SkipList}^i = MBR_{block}^{current}+...+MBR_{block}^{current+\alpha^i}
	\label{MBRSkipList}
\end{equation}

Therefore, the first level's MBR in the skip list can be used to determine whether there are existing transactions that satisfy the query condition in the next $\alpha^i$ blocks. 

\begin{algorithm}[!b]  
\caption{InterQuery($block_{current}, Q$)}
\label{alg:InterQuery}
\hspace*{0.02in}{\bf Input:} Query condition $q = <Q_{discrete}, MBR_{query}>$

\begin{algorithmic}[1]
\FOR{$BF_i and MBR_i$ in $Skiplist$}
    \IF{$BF_i$.isContain() AND $MBR_i$.isIntersect()}
        \IF{i!=0}
            \STATE InterQuery$(block_{current+\alpha^{i}}, Q)$ 
        \ELSE
            \STATE TraverseBlock($block_{current}, block_{current+\alpha^{i}}$)
        \ENDIF
    \ENDIF
\ENDFOR 
\end{algorithmic}
\end{algorithm}

\subsection{Inter-block Query of Discrete-valued Attributes}
For the inter-block query of discrete-valued attributes, we can use bloom-filters in the skip list to help us query quicker. In this way, we also save the time of traversing unnecessary bloom-filters at the head of each block. The specific query algorithm is shown in algorithm \ref{alg:InterQuery}. This can be illustrated by the fact that if $\alpha=2$, we will set the first block of the blockchain as the current block and query the first level's bloom filter in the skip list of the current block. If the return result is true, we will query the bloom-filters $BF_{block}^{2}$, $BF_{block}^{3}$ at the head of the next two blocks. If false is returned, the second level's bloom filter in the skip list of the current block is checked. Eventually, when the $SkipList_{BF}^i$ returns true, it proves that there are might existing blocks that satisfy the query condition between the $({2^{i-1}})^{th}$ block or $({2^{i}})^{th}$ block. Then we need to set the current block as the $({2^{i-1}})^{th}$ block and follow the steps above to continue the query process.

\subsection{Inter-block Query of Continuous-valued Attributes}

Similar to the inter-block query of discrete-valued attributes, we use MBR at the head of each block to generate the $SkipList_{MBR}$ and propose an efficient scheme for the inter-block query of continuous-valued attributes. We set the current block as the first one which denotes as $block_{current}$. Querying process is started from the first level in the skip list of the first block. If the returned result is true, we need to check the MBRs of the next two blocks. If false is returned, the second MBR and MBR behind in the skip list is queried. There are some transactions in the blocks between ${(\alpha^{i-1})}^{th}$ and ${(\alpha^i)}^{th}$ satisfying the query conditions if the check result of the $SkipList_{MBR}^i$ returns true. Then we set the ${(\alpha^{i-1})}^{th}$ block as the current block and continue to query the rest blocks as described above.

\section{Performance analysis}
\subsection{Analysis of the Efficiency}
We analyze the efficiency of our proposed query schemes on the blockchain. Firstly, the intra-block query cost for multi-dimensional continuous-valued attributes is similar to R-tree~\cite{yang2009authenticated}. We use $d_{f}$ and $d_{l}$ to denote the average fan-out of the leaf nodes and the internal nodes in each blocks' R-Tree respectively. In each block, if the number of transactions is $N_{block}$, the number of leaf nodes and internal nodes in this R-Tree is $\frac{N_{block}}{d_{f}}$ and $\frac{N_{block}}{d_{l}}$ ~\cite{theodoridis1996model}. In the unit space $[0,1]^d$ which contains d dimensions, the probability of two rectangles $R_1$ and $R_2$ overlap is as follows in equation (\ref{probabilityOfOverlap}), where $R_{l}^{i}$ express the rectangle R's length along the $i^{th}$ dimension \cite{pagel1993towards} .

\begin{equation}
	P_{overlap} = \displaystyle\prod_{i = 1}^{d}(R_{1,l}^{i}+R_{2,l}^{i}) \label{probabilityOfOverlap}
\end{equation}

We assume that the total sample space is $[0,s^{\frac{1}{d}}]^d$ and the size of each leaf nodes are the same which we denote as $S_{1}=S_{2}=...=S_n$, so the size of each leaf node equals to $\frac{s \cdot d_{f}}{N_{block}}$. Similarly, the size of each internal node in level j is $\frac{s}{d_{n}^j}$. If the length of query condition for continuous-valued attributes in each dimension is $Q^{l,i}_{r}$ and the length for each node in MerkleRB-Tree in every dimension is the same, then the number of nodes in each block's MerkleRB-Tree that needs to be accessed can be computed like in equation (\ref{accessContinuousNodes}).

\begin{equation}
	N_{q} = \displaystyle\prod_{i = 1}^{d}(\sqrt[d]{\frac{s \cdot d_{f}}{N_{block}}}+Q^{l,i}_{r})\cdot d_f + \sum_{j=0}^{h-2} d_l^j \cdot \displaystyle\prod_{i = 1}^{d} (\sqrt[d]{\frac{s}{d_{l}^j}}+Q^{l,i}_{r})
	\label{accessContinuousNodes}
\end{equation}

where the height of the MerkleRB-Tree is $h=1+\log_{d_l}{\frac{s \cdot d_{f}}{N_{block}}}$. Therefore, the total average cost of continuous-valued attributes' query in each block is $C_{range}$, and the cost of querying continuous-valued attributes each node in MerkleRB-Tree is $C_{access}$.
We illustrate it in equation (\ref{totalCost}).

\begin{equation}
	C_{range} = C_{access}\cdot N_{q}
	\label{totalCost}
\end{equation}

For discrete-valued attributes, we assume that the average probability of a bloom filter contains the discrete query condition value $Q_{dis}$ in each block can be shown in equation (\ref{proOfBF}), where the notation $\theta$ expresses the number of times that $Q_{dis}$ appears in the current block. We represent the cost of querying for discrete-valued attributes in equation (\ref{costOfBf}).

\begin{equation}
	P_{BF}(BF, Q_{dis})=\frac{\theta}{N_{block}}
	\label{proOfBF}
\end{equation}

\begin{equation}
	C_{dis} = C_{BF}\cdot(\frac{d_f}{N_{block}}+\sum_{j=0}^{h-2}d_l^j\cdot \frac{d_f \cdot f_n^{h-2-j}}{N_{block}})
	\label{costOfBf}
\end{equation}

When the query condition contains the continuous-valued attributes and discrete-valued attributes together, it is obvious that total cost for query condition contains both discrete-valued attributes and continuous-valued attributes equals to the right hand of the equation (\ref{costTotal}), in which $C_{access}$ denotes the cost of accessing a node in MBR-Tree.

\begin{equation}
	\begin{aligned}
		C_{total} = C_{access}\cdot \frac{d_f}{N_{block}} \cdot \displaystyle\prod_{i = 1}^{d}(\sqrt[d]{\frac{s d_{f}}{N_{block}}}+Q^{l,i}_{r}) \\
		+\sum_{j=0}^{h-2}d_l^j \cdot  \displaystyle\prod_{i = 1}^{d} (\sqrt[d]{\frac{s}{d_{l}^j}}+Q^{l,i}_{r}) \cdot \frac{d_f \cdot f_n^{h-2-j}}{N_{block}}
		\label{costTotal}
	\end{aligned}	
\end{equation}

For the inter-block query, we use a skip list to decrease the query cost for both discrete-valued attributes and continuous-valued attributes. The time complexity of the skip list query is $O(log(n))$ \cite{pugh1990skip}. However, the cost of improving query efficiency is increasing its spatial complexity. And the relationship between time complexity and space complexity in the inter-block query scheme we propose is a negative correlation according to the settings of $\alpha$ and this will be discussed in the next section.

\section{Experiments}
In this part, we implement and test the performance of the inter-block query and intra-block query respectively.

\subsection{Experiment Setting}
\textbf{Dataset} We use a public dataset from kaggle \footnote{https://www.kaggle.com/tejashvi14/employee-future-prediction}. 
In this dataset, it contains different employees with multi-dimensional attributes including continuous value attributes and discrete value attributes. For this experiment, we choose the year of joining company and the age to do the multi-dimensional range query, and choose city as the discrete value. So for each transaction, we can use $Q = < year, age, {city}>$ to represent the query condition.

\textbf{Environments} All the experiments are running a computer which is equipped with Intel Core i7 CPU with 6 cores, 3.2GHz for each core. The memory of the computer is 16GB memory on Window 10 operating system. And the JDK version we use is JDK 1.8. 

\subsection{Performance Evaluation}

\begin{figure}
	\centering
	\includegraphics[scale=0.42]{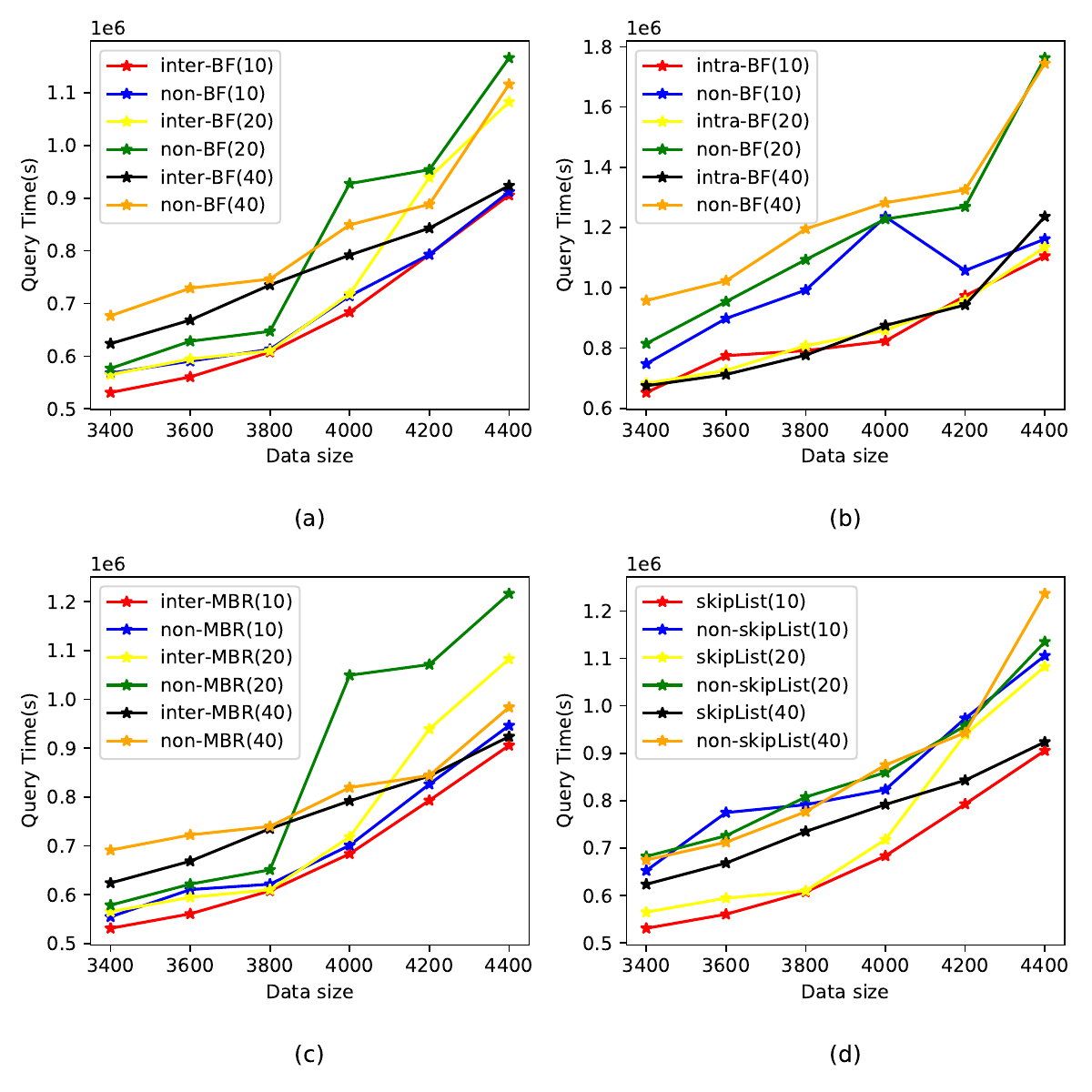}
	\caption{Query Performance}
	\label{fig:score2x2}
\end{figure}

\textbf{Query for Discrete-value Attributes:} To verify the efficiency of the query method we propose above, we test the inter-block and intra-block query performance for the discrete-valued attributes. The results are shown in Figure \ref{fig:score2x2}(a) and Figure \ref{fig:score2x2}(b). 

In figure \ref{fig:score2x2}(a), we test our proposed inter-block query schemes on blockchain with different numbers of transactions from 3400 to 4400. We also put the different number of transactions 10, 20 and 40 in each block. We choose the scheme without $SkipList_{BF}$ as the baseline. We can see that the query time of our schemes are less than the baseline scheme when each block stores the same amount of data. This is because the fact that the method we propose saves time for querying unnecessary blocks when using $SkipList_{BF}$. In addition, with the increasing number of blocks, the efficiency of our plan is more obvious. 

In figure \ref{fig:score2x2}(b), we put different amounts of data 10, 20 and 40 in each block. As for the circumstance which does not have bloom-filters, we cannot quickly exclude subtrees that do not need to be traversed in MBR-Tree through the discrete-valued query condition. Thus, we need to traverse the nodes that only satisfy the continuous-valued query condition and return the correct nodes. By comparing our proposed intra-block query method for the discrete-value attribute with the baseline scheme without bloom-filters, when the amount of data inside the block is the same, it can be seen that the performance of our scheme is better than the baseline scheme. As the amount of data inside the block increases, the query cost of our solution increases smoothly, whereas the query cost of the baseline method increases apparently. The reason lies in that as the number of nodes increases, the baseline method needs to query more useless subtrees, and the method we propose can solve this type of problem effectively. 

\textbf{Query for Continuous-value Attributes:} For query continuous-valued attributes, it is obvious that the intra-block query performance of our scheme using R-Tree is much better than non-index query scheme \cite{theodoridis1996spatio}. Moreover, the performance results of continuous-valued inter-block query efficiency for multi-dimensional data are shown in Figures \ref{fig:score2x2}(c). We choose the method without $SkipList_{MBR}$ as the baseline method. We can see that our scheme performs better than the Baseline scheme, since our proposed scheme for inter-block queries saves the time cost to search unnecessary blocks in blockchain by using $SkipList_{MBR}$. 

In figure \ref{fig:score2x2}(d), we contrast the method without skip list for inter-block query process to our scheme. It can demonstrate the advantages of generating and applying skip list for the inter-block query process in blockchain-based FL.
 
\section{Conclusion}
In this paper, we optimize the query efficiency of selecting participants in blockchain-based FL by modifying the blockchain's structure. By analyzing and comparing the existing query schemes, our scheme that contains intra-block query and inter-block query has superority on query performance. In the future, we will further do explorations in industrial platform of blockchain for varies fields.

\section*{Acknowledgement}
This work is supported by the Sichuan Provincial Key Research and Development Program $($2020YFQ0056, 2021ZHCG0001, 2021YFG0132, 2021GFW046, 2022YFSY0005,  22ZDZX0046$)$ and No.10, Blockchain Incentive Study in Sharing Economy.





\bibliographystyle{ieeetr} 
\bibliography{ref.bib}
\end{document}